\documentclass{conm-p-l}
\usepackage{amssymb,latexsym, amscd}
\usepackage[all]{xy}
\usepackage{graphicx}
 \newtheorem{theorem}{Theorem}[subsection]

 \newtheorem{proposition}[theorem]{Proposition}
 \theoremstyle{definition}
 
 \theoremstyle{definition}
 
 \theoremstyle{remark}
 
 \numberwithin{equation}{subsection}

\usepackage{latexsym}
\usepackage{amssymb}

\newcommand{\ben}{\begin{equation}}
\newcommand{\een}{\end{equation}}

\newcommand{\integer}{\ensuremath{{\mathbb Z}}}

\newcommand{\real}{\ensuremath{{\mathbb R}}}
\newcommand{\complex}{\ensuremath{{\mathbb C}}}

\newcommand{\rational}{\ensuremath{{\mathbb Q}}}

\newcommand{\SL}[1]{\ensuremath{{\mathrm {SL}_{ #1 }}}}

\newcommand{\DD}{{\mathcal D}}
\newcommand{\PP}{{\mathcal P}}

\newcommand{\LL}{\mathcal{L}}
\newcommand{\MM}{\mathcal{M}}
\newcommand{\HH}{\mathcal{H}}

\newcommand{\PPP}{\mathcal{P}}

\newcommand{\Proj}{\mathbf{P}}
\newcommand{\Tate}{\mathbf{L}}

\newcommand{\Aut}{\mathrm{Aut}}
\newcommand{\rank}{\mathrm{rank}}

\newcommand{\Diff}{\mathrm{Diff}}
\newcommand{\charac}{\mathrm{char}}
\newcommand{\Orbit}{\mathrm{Orb}}
\newcommand{\Map}{\mathrm{Map}}
\newcommand{\trace}{\mathrm{Tr}}

\newcommand{\Xx}{\mathsf{X}}
\newcommand{\Ww}{\mathsf{W}}
\newcommand{\Tt}{\mathsf{T}}

\newcommand{\Loop}{\mathsf{L}}
\newcommand{\Orb}{\mathbf{Orb}}
\newcommand{\Obj}{\mathrm{Obj}}
\newcommand{\Mor}{\mathrm{Mor}}
\newcommand{\Spec}{\mathrm{Spec}}
\newcommand{\Bun}{\mathrm{Bun}}
\newcommand{\VAR}{{\mathcal{V}}{{ar}}}

\newcommand{\timests}{\: {}_{t}  \! \times_{s}}

\newcommand{\ord}{{\rm ord}}

\newcommand{\age}[1]{\ensuremath{{\mathrm{age}}{(#1)}}}

\newcommand{\Korb}[1]{\ensuremath{{}^{#1} K_{\mathrm{orb}}}}

\begin{document}

\title
{A Localization Principle for Orbifold Theories}

\author{Tommaso de Fernex, Ernesto Lupercio, Thomas Nevins,  Bernardo Uribe}

\thanks{The first author was partially supported by
a University of Michigan Rackham research grant and the MIUR of
the Italian government, National Research Project ``Geometry on
Algebraic Varieties" (Cofin 2002). The second author was partially
supported by CONACYT-M\'exico. The third author was partially
supported by an NSF Postdoctoral Fellowship at the University of
Michigan. Research of the fourth author was partially carried out
during his visit at the Max Planck Institut, Bonn.}

\thanks{The authors would like to thank A. Adem, L. Borisov, J. Morava, M. Poddar, Y. Ruan and G. Segal for useful conversations during the preparation of this paper.}

\address{Department of Mathematics, University of Michigan,
Ann Arbor, MI 48109, USA}
\email{defernex@umich.edu}
\address{Departamento de Matem\'{a}ticas, CINVESTAV,
     Apartado Postal 14-740
     M\'{e}xico, D.F.  07000 M\'{E}XICO}
\email{lupercio@math.cinvestav.mx}
\address{Department of Mathematics, University of Illinois
at Urbana-Champaign, Urbana, IL 61801, USA}
\email{nevins@math.uiuc.edu}
\address{Department of Mathematics, University of Michigan,
Ann Arbor, MI 48109, USA}
\email{uribeb@umich.edu}

\begin{abstract}
In this  article, written primarily for physicists and geometers,
we survey several manifestations of a general localization
principle for orbifold theories such as $K$-theory, index theory,
motivic integration and elliptic genera.
\end{abstract}

\dedicatory{Dedicated to Prof. Samuel Gitler on the occasion of his 70th birthday.}

\maketitle

\section{Orbifolds}

In this paper we will attempt to explain a general localization
principle that appears frequently under several guises in the
study of orbifolds. We will begin by reminding the reader what we
mean by an orbifold.

The most familiar situation in physics is that of an orbifold of
the type $\Xx=[M/G]$, where $M$ is a smooth manifold and $G$ is a
finite group acting\footnote{We will consider mostly right
actions. Thus, instead of writing $gx$ for the action of $g$ in $x$
we will write  $xg$, the action being $(x,g) \mapsto xg$.}
smoothly on $M$; namely, we  give ourselves a homomorphism $G \to
\Diff(M)$. We make a point of distinguishing the orbifold
$\Xx=[M/G]$ from its quotient space (also called orbit space)
$X=M/G$. As a set, as we know, a point in $X$ is an orbit of the
action: that is, a typical element of $M/G$ is $\Orbit(x)=\{ xg \mid g
\in G \}$.

For us an orbifold $\Xx=[M/G]$ is a smooth category\footnote{
While this may sound slightly far-fetched at first, tolerating this
definition pays off handsomely in simplifying several arguments.}
(actually a topological groupoid) whose objects are the points of $M$, $\Xx_0
= \Obj(\Xx) = M$, and we insist on remembering that $\Xx_0 =
\Obj(\Xx)$ is a \emph{smooth manifold}. The arrows of this
category are $\Xx_1 = \Mor(\Xx)=M\times G$ again thinking of it as
a smooth manifold. A typical arrow in this category is $$ x
\stackrel{(x,g)}{\longrightarrow} xg,$$ and the composition of two
arrows looks like
$$
\xymatrix{
x \ar[r]^{(x,g)} \ar@/_/[rr]_{(x,gh)} & xg \ar[r]^{(xg,h)} & xgh.
}
$$

As we have already pointed out, an important property of this
category is that it is actually a groupoid: indeed, every arrow
$(x,g)$ has an inverse (depending smoothly on $(x,g)$), to wit
$(x,g)^{-1} = (xg, g^{-1})$.

To be fair, the definition of an orbifold is somewhat more
complicated. First, we must impose some technical conditions on
the groupoids that we will be working with. Second,  we must
consider an equivalence relation (usually called \emph{Morita
equivalence}, related to equivalence of categories)
on the family of all smooth groupoids. Then one can roughly say
that an orbifold is an equivalence class of groupoids
\cite{MoerdijkSurveyOrbifolds, LupercioUribeKTheory}. Choosing a
particular groupoid to represent an orbifold is akin to choosing
coordinates for a physical system, and clearly the theories we are
interested in should be invariant under such freedom of choice.

For example, consider the manifold $N = M \times \integer_2$
consisting of two disjoint copies of $M$, and the group $H = G
\times \integer_2$, and let $H$ act on $N$ by the formula $$ (m,
\epsilon_0) \cdot (g,\epsilon_1) = (mg, \epsilon_0 \epsilon_1). $$
Then not only are $N/H\cong M/G$ homeomorphic,  but moreover $\Xx
\cong [N/H] \cong [M/G]$ are equivalent groupoids, while clearly
$N \neq M$ and $H \neq G$.

From now on we will be rather cavalier about this issue and always
choose a groupoid to represent an orbifold. All our final results
will be independent of this choice.

To a groupoid $\Xx$ we will often need to associate a useful
(infinite-dimensional) space, denoted by $B\Xx$, called its
classifying space. We start by drawing a graph, putting a vertex
for every object $m\in\Xx_0$ (remembering the topology of the
space of objects). Then we draw an edge for every arrow
$\alpha\in\Xx_1$. We fill in with a 2-simplex $\Delta^2 =
\{(x,y,z) | x+y+z=1,\ x,y,z\geq0\}$ every commutative triangle in
the graph (with edges $(\alpha,\beta,\alpha\circ\beta)$). Then we
fill with a 3-simplex any commutative tetrahedron, and so on (see
\cite{SegalClassifyingSpaces}). There is always a canonical
projection map $\pi_\Xx \colon B\Xx \to X$ so that
$\pi_\Xx^{-1}(\Orbit (x)) \simeq BG_x$, where $G_x$ is the
stabilizer of $x$, that is, the subgroup of $G$ fixing $x$.

For example, if the orbifold in question is $\Xx=[*/G]$, a group
acting on a single point, then $B\Xx = BG$ is the classifying
space of $G$. This space has the property that the set of
isomorphism classes of principal $G$-bundles $\Bun_G(Y)$ over $Y$
is isomorphic to $[Y,BG]$, the set of homotopy classes of
continuous maps from $Y$ to $BG$. When $\Xx=[M/G]$, it is customary
to write $B\Xx=M\times_G EG$, also known as the \emph{Borel
construction of the group action $M\times G \to M$}. For a general
groupoid $\Xx$, we refer the reader to
\cite{MoerdijkClassifyingToposBook} for a description of what
exactly $B\Xx$ classifies. It is important to remark that $B\Xx$
depends on which representation one considers for $\Xx$, but its
\emph{homotopy type} is independent of this choice.

A final remark: there are orbifolds $\Xx$ that \emph{cannot} be
represented by a groupoid of the form $[M/G]$. In other words, in
spite of the fact that there is indeed a groupoid representing
$\Xx$, nevertheless there is no manifold $M$ with a \emph{finite}
group action $G$ so that $\Xx\cong [M/G]$. We say in this
situation that the orbifold in question is \emph{not} a
\emph{global quotient}. Examples are given by the toric
orbifolds $\Ww(a_0,\dots,a_n)$ whose quotient spaces are
the weighted projective spaces $\Proj(a_0,\dots,a_n)$
(here $a_i$ are coprime positive integers). For simplicity, let us
discuss the case of the orbifold
$\Ww(1,2)$ whose quotient space is the weighted projective line
$\Proj(1,2) \cong \Proj^1$. One way to describe $\Ww(1,2)$ is through the
system of local charts:
$$\xymatrix{
 & [\complex^\times/\{1\}] \ar[dl]_{z \mapsto {1}/{z^2}} \ar[dr]^{z\mapsto z} & \\
 [\complex/\integer_2] & & [\complex/\{1\}].
}$$
If $\Ww(1,2)$ were Morita equivalent to a groupoid $[M/G]$, then
this would induce a homomorphism $\rho \colon G \to \integer_2$
(this follows by looking at the unique point in $\Ww(1,2)$ with
isotropy $\integer_2$). Therefore the orbifold $[M'/\integer_2]$
with $M' :=M/ker(\rho)$ would be equivalent to $\Ww(1,2)$. But
this is a contradiction because any action of $\integer_2$ in a
compact surface cannot have only one fixed point.

This example might be a source of misunderstanding
because weighted projective spaces are indeed quotient
varieties of manifolds by actions of finite groups.
For instance, in our example, $\Proj(1,2)$
is isomorphic to the quotient of $\Proj^1$ by
$\integer / 2 \integer$ under the action $[x,y] \mapsto [x,-y]$ in homogeneous
coordinates. On the other hand, although the orbifold $\Ww(1,2)$
can be presented as a quotient of a manifold by an action of a Lie
group, namely $[\complex^2-\{0\} / \complex^\times]$ with
$\lambda\cdot (x,y) \mapsto (\lambda^2 x, \lambda y)$, it is not equivalent to
global quotient by a finite group. It is worth pointing
out that it is still an open question whether every compact orbifold
can be presented (up to Morita equivalence) as the quotient of a
manifold by a Lie group \cite{henriques}.

\section{Orbifold $K$-Theory}

In their seminal paper \cite{DHVW}, Dixon, Harvey, Vafa,
 and Witten  defined the
orbifold Euler characteristic of an orbifold $\Xx=[M/G]$ by the
formula
\begin{equation}\label{EulerCharOrbI}
\chi_\Orb(\Xx) = \frac{1}{|G|} \sum_{gh=hg} \chi(M^{g,h}),
\end{equation}
where $(g,h)$ runs through all the pairs of commuting elements of
$G$ and $M^{g,h}$ is the set of points in $M$ that are fixed both
by $g$ and by $h$. They obtained this formula by considering a
supersymmetric string sigma model on the target space $M/G$ and
noting that in the known case in which $G=\{1\}$ the Euler
characteristic of $\Xx=M$ is a limiting case (over the worldsheet
metric) of the partition function on the 2-dimensional torus.

In essentially every interesting example, the \emph{stringy
orbifold Euler characteristic} $\chi_\Orb(\Xx)$ is not equal to
the ordinary Euler characteristic of the quotient space $\chi(X)$.
More interestingly, $\chi_\Orb(\Xx)$ is truly independent of the
particular groupoid representation, namely if
$\Xx=[M/G]\cong[N/H]$ then it doesn't matter which representation
one uses to compute $\chi_\Orb(\Xx)$. In other words, this is a
truly physical quantity independent of the choice of coordinates.
This last remark, which can be readily verified by the reader, is
quite telling, since \emph{a priori} the sigma model depends on the
particular groupoid representation. But as the theory is indeed
physical, the final partition function is independent of the choice
of coordinates.

Moreover, since the partition function of the theory is
physical, one may expect a stronger sort of invariance. Should
there be a well-behaved (smooth) resolution of $X$ defining the
same quantum theory, then one should have that the Euler
characteristic of the resolution is the same as that of the
original orbifold.  Here we are shifting our point of view,
thinking of an orbifold as the quotient  space with a mild type of
singularities. It is a remarkable fact in algebraic geometry \cite{CartanQuotient} that
in good cases, remembering $X$ plus some additional algebraic data
(for example the structure sheaf), one can recover $\Xx$. This
point of view has proved extremely fruitful as we shall see. In
any case, it often happens that there are
 resolutions of $X$, the \emph{crepant resolutions},
 for which the quantum theory is the same as
that for $X$. We will come back to this
later.

There is, of course, a far more classical interpretation of the
Euler characteristic, the topological interpretation.  The classical
interpretation of the Euler characteristic in terms of
triangulations tells us that the Euler characteristic is
 the alternating sum of the Betti
numbers, namely, the ranks of the cohomologies of the space in
question. Thus, a natural question is whether there is a cohomology
theory for an orbifold that is physical and that simultaneously
produces the appropriate Euler characteristic
\eqref{EulerCharOrbI}. One is first tempted to consider
equivariant cohomology $H^*_G(M) = H^*(M \times_G EG)$ but
unfortunately the relation between cohomology and Euler
characteristic breaks down, for the expression
\eqref{EulerCharOrbI} is not recovered.

Considering the orbifold $\Xx=[*/G]$ consisting of a finite group
acting on a single point gives us a clue into the right answer. In
this case, $\chi_\Orb([*/G])$ becomes the number of pairs of
commuting elements in $G$ divided by $|G|$. An amusing exercise in
finite group theory readily shows that this is the same as the
number of conjugacy classes of elements in $G$. Given a finite
group there are two basic quantities that we can consider, its
group cohomology $H^*(BG)$ and its representation ring $R(G)$.
While equivariant cohomology is akin to group cohomology, it is
\emph{equivariant $K$-theory} $K_G(M)$ that is intimately
related to representation theory. For a start, $K_G(*)=R(G)$.

As a first test, we consider an orbifold $\Xx=[M/G]\cong[N/H]$
and see whether the theory is invariant under the representation.
This is not too hard (see for example \cite{LupercioUribeKTheory,
AdemRuan}), and hence it fully deserves the name of \emph{orbifold
$K$-theory} and can unambiguously be written as $K_\Orb(\Xx)=K_G(M)
\cong K_H(M)$.

The second test is to see whether we can recover Formula
\eqref{EulerCharOrbI}. That this is possible was first observed by
Atiyah and Segal \cite{AtiyahSegal1}. The idea is to use the Segal
character of an equivariant vector bundle. Let us remember that
the basic cocycles of equivariant $K$-theory are $G$-equivariant
vector bundles \cite{AtiyahKtheory}, namely bundles $p \colon E
\to M$ over the $G$-manifold $M$ with a $G$-action by bundle
automorphisms on all of $E$ that extends the action on $M$
(considered as the zero section) and that is fiberwise linear.
Should there be a fixed point $m\in M$, then $E_m := p^{-1}(m)$
becomes a representation of $G$; in particular, if the space $M$ is
a point then a $G$-equivariant vector bundle over $M$ is the
same as a representation of $G$ (by choosing a basis we get a
matrix for every $g\in G$).

The (Segal) character of an equivariant vector bundle is an
\emph{isomorphism} \cite{Segalunpublished,
MoerdijkSurveyOrbifolds} of the form
\begin{equation} \label{SegalIso}
K_G(M) \otimes \complex \stackrel{\cong}{\longrightarrow} \bigoplus_{(g)} K(M^g)^{C(g)} \otimes \complex,
\end{equation}
where the sum is over all conjugacy classes $(g)$ of elements $g\in G$.

The character isomorphism is  explicitly given by the expression
 \begin{eqnarray*}
K_G(M) \otimes \complex & \to &K(M^g)^{C(g)} \otimes \complex\\
 E \otimes 1& \mapsto& \charac(E)(g) = \sum_\zeta (E|_{M^g})_\zeta \otimes \zeta.
 \end{eqnarray*}
Here the sum is over all roots of unity $\zeta$, the symbol
 $(\;\;)_\zeta$
denotes the $\zeta$-eigenspace of $g$, and finally $M^g$ is the
subspace of fixed point under $g$ of $M$. We call this isomorphism
the \emph{Segal localization formula} (for it localizes
equivariant $K$-theory to ordinary $K$-theory of the fixed point
sets).\footnote{Remarkably enough this is indeed related to the
localization of equivariant $K$-theory as an $R(G)$-module with
respect to prime ideals \cite{SegalEquivariantK}.}  Clearly, in the
case in which $M$ is a point, this recovers the usual theory of
characters for the finite-dimensional representations of a finite
group.

From Segal's isomorphism \eqref{SegalIso} we conclude immediately
that \cite{AtiyahSegal1, ConnesBaumes, UribeThesis} $$ \rank
K_G^0(M) - \rank K_G^1(M) = \sum_{(g)} \chi(M^g/C(g)) =
\frac{1}{|G|} \sum_{gh=hg} \chi(M^{g,h}) = \chi_\Orb(\Xx). $$
Here we have applied
the algebraic equality $$
\chi_\Orb(\Xx) = \sum_{(g)} \chi(M^g/C(g)),$$
which follows by an inclusion-exclusion
argument \cite{HH}; in the next
section we talk about a geometric explanation for this algebraic
fact.

For now let us mention that the theory described in this section
can be generalized to orbifolds that are not necessarily global
quotients \cite{LupercioUribeKTheory, AdemRuan}. This is done as
follows. We will denote by $\Xx_0$ and $\Xx_1$ the set of objects
and morphism of our groupoid respectively, and the structure maps
by
$$\xymatrix{
       \Xx_1 \timests \Xx_1 \ar[r]^{\ \ \ m} & \Xx_1 \ar[r]^i &
       \Xx_1 \ar@<.5ex>[r]^s \ar@<-.5ex>[r]_t & \Xx_0 \ar[r]^e & \Xx_1,
}$$
where $\Xx_1 \timests \Xx_1$ is the subspace of $\Xx_1 \times
\Xx_1$ such that whenever $(\alpha,\beta) \in \Xx_1 \timests
\Xx_1$ then the target of $\alpha$ equals the source $\beta$;  $s$
and $t$ are the source and the target maps on morphisms, $m$ is
the composition arrows, $i$ gives us the inverse morphism, and $e$
assigns the identity arrow to every object.

We define a \emph{vector orbibundle} over $\Xx$ to be a pair
$(E,\tau)$ where $E$ is an ordinary vector bundle over $\Xx_0$ and
$\tau \colon s^* E \stackrel{\cong}{\longrightarrow} t^* E$ is an
isomorphism of vector bundles over $\Xx_1$.

The set of isomorphism classes of such orbibundles is denoted by
$Orbvect(\Xx)$ and its Grothendieck group by $\Korb{}^0(\Xx)$
\cite{LupercioUribeKTheory}.

This coincides with equivariant $K$-theory if the
orbifold happens to be of the form $[M/G]$.

\section{Loop Orbifolds}

Let us consider now a Riemannian metric on $M$. There is then a
family of canonically defined operators: the Laplacians on
$k$-forms $\Delta^k$. These are related to a quantum field theory
whose fields are maps from intervals the circle to $M$. Roughly
speaking, the Lagrangian of the theory is given by
$$ L(\phi) = \frac{1}{2} \int | d{\phi} |^2.$$
All the information of such quantum theory is contained in the
spectrum of the Laplacian.
Recovering the classical theory from the quantum one is
``hearing the shape of the drum.''
In any case, the Feynman functional integration approach for the
theory allows us to compute an integral over the free loop space
of the manifold $\LL(M) = \Map(S^1;M)$ by stationary phase
approximation as an integral over $M$.


This quantum field formalism is related to the heat equation
$$ \partial_t \omega + \Delta^k \omega=0, $$
whose solution is given by the heat flow $e^{-t \Delta^k}$. In particular
the fundamental solution for the trace of the heat kernels is given by
$$\sum (-1)^k \trace(e^{-t \Delta^k}) = \int_{\LL M} e^{t^{-1} L(\phi)} \DD \phi,$$
where $\DD \phi$ is the formal part of the Wiener measure on $\LL M$.

It turns out that the the sum of the traces of the heat kernels is
independent of $t$. The long time limit of this sum equals the
Euler characteristic (by recalling Hodge's theorem, which
identifies the $k$-th Betti number of $M$ as the dimension of the
kernel of $\Delta^k$), and the short time behaviour is given by an
integral of a complicated curvature expression.

 If the dimension
of the manifold is 2, this equality of long and short time
behaviour of the heat flow leads to the Gauss-Bonet theorem
\begin{equation} \label{GaussBonnet}
\int_M K dA = \chi(M),
\end{equation}
where $K$ is the Gaussian curvature and $dA$ is the volume element.


In fact we have oversimplified: we can do better than to simply
recover the Euler characteristic. Suppose that $M$ is a spin
manifold;
 then we can recover by this procedure
the \emph{index of the Dirac operator}. We will come back to this
observation in the next section. But before we do that let us see
how we stand in the orbifold case.

To try to apply these methods to an orbifold $\Xx$ (replacing the
r\^ole of $M$ above), we must be able to say what is the candidate
to replace $\LL M$. This was done for a general orbifold in
\cite{LupercioUribeLoopGroupoid}.

The idea is that to a groupoid $\Xx$ we must assign a new
(infinite-dimensional) groupoid $\Loop \Xx$ that takes the place
of the free loopspace of $M$. The construction in
\cite{LupercioUribeLoopGroupoid} furthermore commutes with the
functor $B$ from groupoids to spaces defined in the first section,
as shown in \cite{LupercioUribeXicotencatl},
in the sense that there is an homotopy equivalence $$ B\Loop \Xx \simeq \LL B
\Xx.$$ In the case in which $\Xx=[M/G]$, we proved that $\Loop \Xx$
admits a very concrete model defined as follows.

The objects of the loop groupoid are given by $$(\Loop \Xx)_0 :=
\bigsqcup_{g \in G} \PPP_g,$$ where $\PPP_g$ is the set of all
pairs $(\gamma,g)$ with  $\gamma :\real \to X$ and $g \in G$ with
$\gamma(t)g = \gamma(2\pi + t)$.

The space of arrows of the loop groupoid is $$(\Loop \Xx)_1 :=
\bigsqcup_{g \in G} \PPP_g  \times G,$$ and the action of $G$ in
$\PPP_g$ is by translation in the first coordinate and conjugation
in the second; that is, a typical arrow in the loop groupoid looks
like
$$(\gamma,g) \stackrel{((\gamma,g);h)}{\longrightarrow} (\gamma \cdot h, h^{-1} gh),$$
or pictorially:

\begin{center}
\includegraphics[width= 3.5 in, height= 1.3 in]{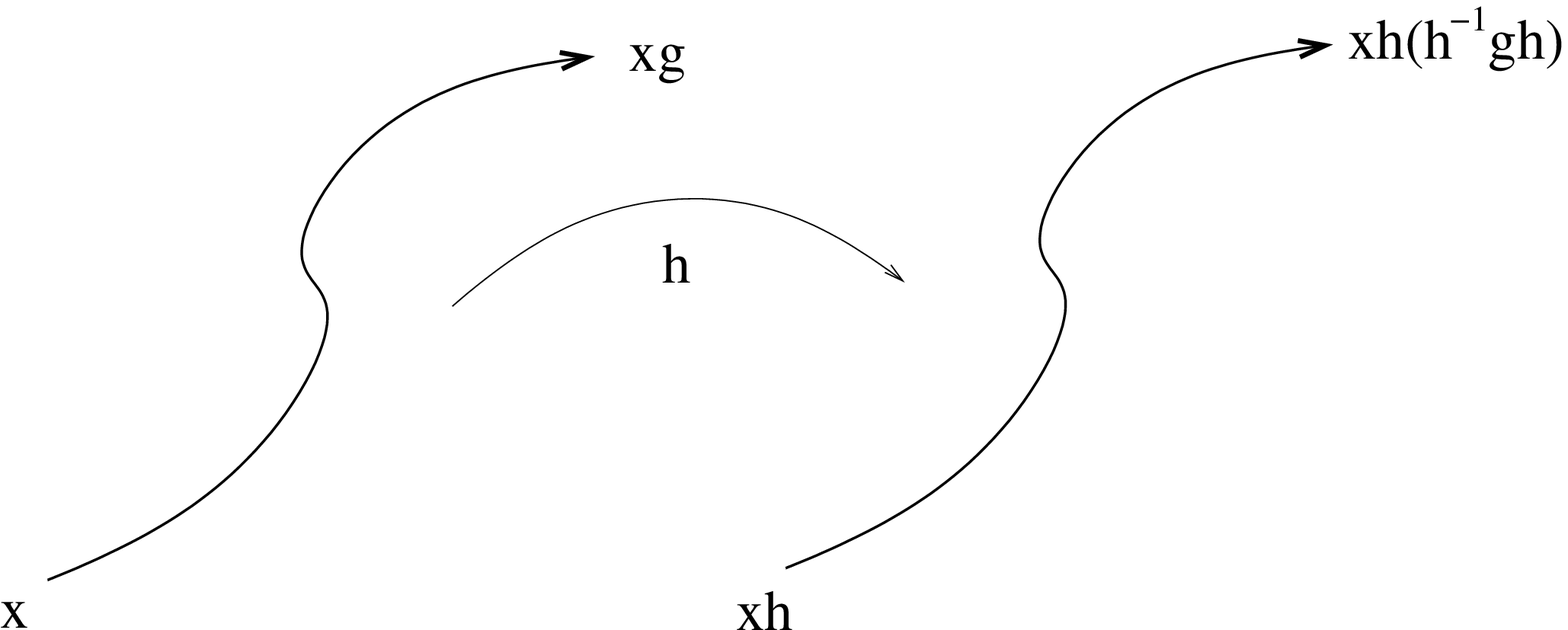}
\end{center}

We are ready to state the basic localization principle.

\begin{theorem}[The Localization Principle \cite{LupercioUribeLoopGroupoid}]
Let $\Xx$ be an orbifold and $\Loop \Xx$ its loop orbifold. Then the
fixed orbifold under the natural circle action by rotation of loops is
\begin{equation}\label{LocalizationPrinciple}
(\Loop\Xx)^{S^1} = I(\Xx),
\end{equation}
where the groupoid $I(\Xx)$ has as its space of objects
$$ I(\Xx)_0 = \{ \alpha \in \Xx_1 \colon s(\alpha) = t(\alpha) \} = \coprod_{m \in \Xx_0} \Aut_\Xx (m)$$ and its space of arrows is $$I(\Xx)_1 = Z(I(\Xx_0))=\{ g\in \Xx_1 \colon \alpha \in I(\Xx)_0 \Rightarrow g^{-1} \alpha g \in I(\Xx)_0 \};$$
a typical arrow in $I(\Xx)$ from
$\alpha_0$ to $\alpha_1$ looks like
$$
        \xymatrix{
        \circ \ar@(ul,dl)[]|{\alpha_0} \ar@/^/[rr]|{g}
        &&\circ \ar@(dr,ur)[]|{\alpha_1^{-1}} \ar@/^/[ll]|{g^{-1}}.
}$$
\end{theorem}

While for a smooth manifold we have $$M=(\LL M )^{S^1},$$ we have,
by contrast, $$\Xx \subset I(\Xx) = (\Loop\Xx)^{S^1},$$; so we
expect the Euler characteristic, $K$-theory, and so on to localize in
$I(\Xx)$ rather than in $\Xx$. The orbifold $I(\Xx)$ is called in
the mathematical literature the \emph{inertia orbifold} of $\Xx$,
and it is, as Chen and Ruan \cite{ChenRuan1} have pointed out (and
as is reflected in their terminology), the classical geometrical
manifestation of the \emph{twisted sectors} of orbifold string
theory \cite{DHVW}.

Indeed, we have that for a general orbifold
$$ \chi_\Orb(\Xx) = \chi(I(\Xx)) $$
and
$$ \Korb{}^*(\Xx)\otimes\complex \cong K^*(I(\Xx))\otimes\complex.$$

For example, in the case of a global quotient $\Xx=[M/G]$, one
can readily verify that
\begin{equation}\label{InertiaGlobal}
I(\Xx) = \coprod_{(g)} [M^g/C(g)],
\end{equation}
recovering thus Segal's localization formula and the orbifold
Euler characteristic of the previous section.

In \cite{LupercioUribeGhost}, the \emph{ghost loop
space} $\LL_s B\Xx$ is defined
 as the subspace of elements $\gamma \in \LL B
\Xx$ so that the composition with the canonical projection
$\pi_\Xx \colon B\Xx \to X$, $\pi_\Xx \circ \gamma$ is constant.
That paper proves the homotopy equivalence
$$B I(\Xx) \simeq \LL_s B \Xx.$$

\section{The McKay Correspondence}

Let us consider now a classical example. Let $G$ be a finite
subgroup of $\SL{2}(\complex)$; then $X=\complex^2 / G$ is called
a \emph{Kleinian quotient singularity}; see \cite{Slodowy, HarpeSiegfried} for
more details and historical discussion.  In the second half of the
19th century, Klein
classified the possible groups $G$ as either cyclic, dihedral or
binary dihedral and gave equations for these singularities in
$\complex^3$.  Let us consider the simplest case in which
$G\cong \integer/r\integer$. We can realize $X$ as a subvariety of
$\complex^3$ by $$ X\colon z^r = xy$$ or, in parametric form,
\begin{equation}
   \begin{array}{c}
      x = u^r \\
      y = v^r \\
      z=uv\
   \end{array}
\end{equation}
as the image of a map $\complex^2 \to \complex^3$ by $G$-invariant
polynomials. We can resolve the singularity
very easily in this case by taking
$(r-1)$-blow ups to obtain $$Y \stackrel{\phi}{\longrightarrow}
X$$ where the exceptional divisor is
$$\phi^{-1}(0)= E_1 \cup E_2 \cup \dots \cup E_{r-1}$$
whose incidence graph is $A_{r-1}$.

On the other hand, $G$ clearly has $r-1$ nontrivial irreducible
representations.

The \emph{McKay correspondence} establishes (among other things) a
one-to-one correspondence between the number of components of the
exceptional divisor in a minimal resolution of the singularity
and the number of nontrivial irreducible
representations of $G$. Notice that in our example this is
equivalent to the statement that \emph{the orbifold Euler
characteristic of $\Xx$ is the same as the ordinary Euler
characteristic of $Y$}. So one may expect that some functional
integral argument may be provided to prove the McKay
correspondence.

There is in fact a rigorous version of the functional integration method
in algebraic geometry discovered by Kontsevich
\cite{KonsevichFamousTalk} and known as \emph{motivic integration}.
We now briefly outline the construction of this method.

Given a smooth complex variety $Y$,
one can define its {\it arc space} $JY$. This
is a scheme whose $\complex$-points
are arcs $\gamma \colon \Spec(\complex[[t]]) \to Y$. The scheme $JY$ is obtained
as the inverse limit of the {\it jet schemes} $J_mY$,
whose $\complex$-points are jets $\gamma_m \colon
\Spec(\complex[t]/(t^{m+1})) \to Y$. The morphisms
$J_pY \to J_mY$, for $0 \le m \le p \le \infty$, are given by truncation.
For any effective divisor $D \subset Y$, one can define
an order function
$$
\ord_D : JY \to \integer_{\ge 0 } \cup \{\infty\},
$$
which to each arc $\gamma$ associates its order of contact $\ord_D(\gamma)$
along $D$. The idea is then
 to ``integrate these functions,'' in some reasonable sense.
But first one needs to introduce the algebra of measurable sets and
the measure. The first is easily defined as the algebra generated
by {\it cylinder sets} in $JY$, namely, inverse images of constructible
sets on finite levels $J_mY$. The measure will then take values in
the so-called {\it motivic ring}.

The motivic ring is constructed as follows: we
fix a complex variety $X$ and assume that $Y$ is an $X$-variety
(that is, a complex algebraic variety of finite type over $X$).
Let $K_0(\VAR_X)$ be the ring generated by $X$-isomorphism classes of
$X$-varieties  subjected to the relation $$\{ V \} = \{ V
\setminus W \} + \{W\}$$
whenever $W$ is a closed variety of a $X$-variety $V$.
The product is defined by
$$\{V\}\cdot\{W\} = \{V\times_X W\}.$$
The zero of this ring is $\{\varnothing\}$, and
the identity is $\{X\}$. We let
$$\MM_X=K_0(\VAR_X)[\Tate_X^{-1}],$$ where $\Tate_X$ is the
class of the affine line over $X$. Finally the motivic ring is the
completion $\hat{\MM_X}$ of $\MM_X$ under a certain natural
dimension filtration \cite{Looijenga,DenefLoeser,deFxLuNevUri}.

Via composition, every subvariety of the jet schemes of $Y$ can be
viewed as an $X$-variety. Thus, one can define the \emph{motivic measure}
of a cylinder $C \subseteq JY$ by fixing a large enough integer $m$
such that $C$ is the inverse image of a constructible set $C_m \subseteq J_mY$
and then setting
$$\mu(C) = \{ C_m \} \cdot \Tate_X^{-m\dim Y} \in \hat{\MM}_X.$$
Then, by suitable stratification, one defines the {\it motivic integral}
$$
\int_{JY} \Tate_X^{-\ord_D} d\mu \in \hat{\MM}_X.
$$
For instance, if $D = \sum a_jD_j$ is a simple normal crossing divisor and we define
$$
D_J^\circ
= \bigcap_{j\in J} D_j \setminus\bigcup_{i \not \in J} D_i,$$
then one has
$$
\int_{JY} \Tate_X^{-\ord_D} d\mu  = \sum_{J\subseteq I} \{D_J^\circ\}
\prod_{j\in J} \frac{\Tate_X -1}{\Tate_X^{a_j+1}-1}.
$$
The power of this theory is a {\it change of variable formula};
this allows us to reduce to computing integrals for divisors with simple
normal crossings (hence apply the above formula)
by replacing any effective divisor $D$ on $Y$
by $D' = K_{Y'/Y} + g^*D$, where $g : Y' \to Y$ is a simple normal crossing resolution
of the pair $(Y,D)$. The theory can be also extended to
singular varieties (under suitable conditions): in this
case the measure itself needs to be opportunely ``twisted''
to make the change of variable formula work.   The resulting
measure is called {\it Gorenstein measure} and denoted by $\mu^{Gor}$.

We can now review the
{\it motivic McKay correspondence} \cite{DenefLoeser, Looijenga, MilesReid}.
To give a formulation of this correspondence that better
fits in the ``localization'' context of this paper, we need to
further quotient the ring $K_0(\VAR_X)$ by identifying
$X$-varieties that become isomorphic after some \'etale
 base change $X'_k \to X_k \subseteq X$
of each piece $X_k$ of a suitable stratification $X = \bigsqcup X_k$
of $X$.
We obtain in this way a new ring: $K_0(\VAR_X)^{et}$.
This leads to the definition of a different motivic ring, which
we denote by $\hat{\MM}_X^{et}$ (the reader will notice that,
if $X$ is a point, then we are not changing anything).

Let $\Xx=[M/G]$, where $M$ is a quasiprojective variety and $G$
is a finite group, let $X = M/G$, and assume that $X$ is Gorenstein.
We can find a resolution of singularities $Y \to X$ with
relative canonical divisor $K_{Y/X}$ having simple normal crossings.
Write $K_{Y/X} = \sum a_j D_j$. Then
the McKay correspondence is given by the identity
\begin{equation}\label{MotivicMcKay}
\sum_{J\subseteq I} \{D_J^\circ\} \prod_{j\in J} \frac{\Tate_X -1}{\Tate_X^{a_j+1}-1}
= \sum_{(g)} \{M^g/C(g)\} \Tate_X^{w(g)} \quad\text{in $\hat{\MM}_X^{et}$},
\end{equation}
where the sum in the left side runs over
conjugacy classes $(g)$ in $G$ and
$w(g)$ are integers depending on the local action
of $g$ on the normal bundle of $M^g$ in $M$.

For instance, by noticing that the Euler characteristic defines a
ring homomorphism $$ \chi \colon K_0(\VAR_X)^{et} \to \integer,$$ it is
easy to see that Formula \eqref{MotivicMcKay} implies the classical
McKay  correspondence,\footnote{Here we are referring only to the
counting statement, and not that we recover the full incidence graph of
$\phi^{-1}(0)$ from the representation theory of $G$, as the
classical correspondence establishes.} which in particular
says that the orbifold Euler characteristic is equal to the
ordinary Euler characteristic of the resolution if the latter
is crepant.

The proof of Formula~\eqref{MotivicMcKay} breaks into three parts.
By the change of variable formula, one has
$$
\int_{JX} \Tate_X^{0} d\mu^{Gor} =
\sum_{J\subseteq I} \{D_J^\circ\} \prod_{j\in J} \frac{\Tate -1}{\Tate^{a_j+1}-1}
\quad\text{in $\hat{\MM}_X$}.
$$
Then, by an accurate study of lifts of the arcs of $X$ to arcs on $M$,
one proves that
$$
\int_{JX} \Tate_X^{0} d\mu^{Gor} =
\sum_{(H)} \{X^H \} \sum_{(h)} \Tate_X^{w(h)} \quad\text{in $\hat{\MM}_X$}.
$$
Here the first sum runs over conjugacy classes $(H)$ of subgroups
of $G$, $X^H \subseteq X$ is the image of the set of points on $M$
whose stabilizer is $H$, and the last sum is taken over conjugacy classes
in $H$. The above identity is the core of the proof.
 Finally, one shows that
$$
\sum_{(H)} \{X^H \} \sum_{(h)} \Tate_X^{w(h)} =
\sum_{(g)} \{M^g/C(g)\} \Tate_X^{w(g)} \quad\text{in $\hat{\MM}_X^{et}$}.
$$
Here is where we need to pass to the ring $\hat{\MM}_X^{et}$.
This last part can be easily verified using certain properties of Deligne-Mumford stacks
(see \cite{deFxLuNevUri}). In general, if we do not perform
the additional localization in the relative motivic ring, but instead work
with the ring $\hat{\MM}_X$,
we do not expect the last identity to hold.

These results have been extended to general
(not necessarily global quotient) orbifolds independently by Yasuda
\cite{Yasuda} and by Lupercio-Poddar \cite{LupercioPoddar}.

In \cite{deFxLuNevUri}, we used a natural homomorphism
from $K_0(\VAR_X)$ to the ring of constructible functions $F(X)$ on $X$
to associate to any motivic integral an element in $F(X)_\rational$,
that is, a rational-valued constructible function on $X$.
In fact, one observes that this construction factors through
$K_0(\VAR_X)^{et}$.\footnote{In particular, this tells us
that the identification performed to define $\hat{\MM}_X^{et}$
does not trivialize the ring too much, as we can still recover
all the information in $F(X)$.}
The result is the following localization
formula for constructible functions:
\begin{equation}\label{constructible-fn}
\sum_{J\subseteq I} \frac{(f|_{D_J^\circ})_* {\bf 1}_{D_J^\circ}}{\prod_{j\in J}(a_j+1)}
= \sum_{(g)} (\pi_g)_* {\bf 1}_{M^g/C(g)} \quad\text{in $F(X)$},
\end{equation}
where $\pi_g : M^g/C(g) \to X$ is the morphism naturally induced
by the quotient map $\pi : M \to X$ \cite[Theorem~6.1]{deFxLuNevUri}.

Motivic integration was used in \cite{deFxLuNevUri}
to define the {\it stringy Chern class} $c_{\rm str}(X)$
of $X$. In the case at hand,
we use the MacPherson transformation \cite{MacPherson} to
deduce from \eqref{constructible-fn} the following localization
formula for the stringy Chern class of a quotient \cite[Theorem~6.3]{deFxLuNevUri}:
$$
c_{\rm str}(X) = \sum_{(g)} (\pi_g)_* c_{\rm SM}(M^g/C(g)) \quad\text{in $A_*(X)$},
$$
where $c_{\rm SM}(M^g/C(g))$ is the Chern-Schwartz-MacPherson class of
$M^g/C(g)$ \cite{MacPherson}. This generalizes and implies Batyrev's formula
for the Euler characteristic \cite{Batyrev}.

\section{Index Theory}

Now the situation is as follows. Suppose that we have a compact
symplectic $2m$-dimensional manifold $N$ with symplectic form $\omega$
  and that $H\colon N \to \real$ is the Hamiltonian function of a
Hamiltonian circle action. Let $F_\alpha$ be the critical
manifolds of $H$ (namely the fixed points of the action) with
critical values $H_\alpha$. The Liouville volume form on $N$ is
${\omega^m}/{m!}$. The Duistermaat-Heckman formula reads
\cite{AtiyahBottMomentMap,DuistermaatHeckman}
$$\int_N e^{\hbar H} \frac{\omega^m}{m!} = \sum_\alpha e^{\hbar
H_\alpha} \int_{F_\alpha} \frac{e^\omega}{E_\alpha},$$ where
$E_\alpha$ is the equivariant Euler class of the normal bundle of
$F_\alpha$ in $N$. If $\hbar$ is taken as purely imaginary, the
integral over $N$ is oscillatory, the submanifolds $F_\alpha$ are
the stationary points of $H$, and   the right-hand side of this
formula is given by stationary phase approximation.

Witten \cite{AtiyahAsterisqueCircularSymm} had the idea of using
the Duistermaat-Heckman formula in the case $N=\LL M$, the free
loop space of a manifold $M$, with Hamiltonian
$$ H(\gamma) = \frac{1}{2} \oint_{S^1} |\gamma'(t)|^2 dt. $$

In this case Atiyah defines a symplectic form on $\LL M$ whenever
$M$ is compact and orientable. Then he goes on to show that \emph{
when $M$ is a Spin manifold, $\LL M$ is orientable}. Moreover, he
shows that the left-hand side of the corresponding Duistermaat-Heckman formula is the
heat kernel expression for the index of the Dirac operator while
the right-hand side is the $\hat{A}$-genus, thus giving the
Atiyah-Singer index theorem.

We do the same now for the loop groupoid. In order to simplify the
calculation, we will consider the case of a global quotient
$\Xx=[M/G]$, but everything that we will say generalizes to
general (non--global-quotient)
 orbifolds. We will suppose thus that $M$ is a compact,
even-dimensional spin manifold such that for every $g \in G$ the
map  $g: M \to M$ given by the action is a spin-structure-preserving
isometry.
We will argue that applying stationary
phase approximation to the integral\footnote{In
\cite{AtiyahAsterisqueCircularSymm} it is explained how to make sense
of this integral.}
\begin{eqnarray}
\int_{\PPP_g} e^{-t E(\phi)} \{{\rm Tr} \ S^+(T_\phi) -{\rm Tr} \ S^-(T_\phi)\} \DD \phi \label{path integral}
\end{eqnarray}
one obtains
$$Spin(M,g) :=ind_g(D^+) = {\rm tr}(g|_{ker D^+}) - {\rm tr}(g|_{coker D^+}),$$ the
value of the $g$-index of the Dirac operator $D^+$ over $M$. Here
$E$ is the energy of the path (Hamiltonian)
$$E(\phi):= \frac{1}{2} \int_0^{2\pi} |\phi'(t)|^2dt,$$
$\DD \phi$ denotes the formal part of the Weiner measure on
$\PPP_g$,  $T_\phi$ is the tangent space at $\phi \in \PPP_g$, and
$S^+, S^-$ denote the two half-spin representations of $Spin(2m)$
($2m =\dim M$).

The real numbers act on $\PPP_g$ by shifting the path
\begin{eqnarray*}
\PPP_g \times \real & \to & \PPP_g\\
(f,s) &\mapsto & f_s : \real \to M\\
 & & f_s(t) := f(t-s)
\end{eqnarray*}
and the fixed point set of this action on $\PPP_g$ consists of the
constant maps to $M^g$ (the fixed point set of the action of $g$
in $M$), that is,
$$(\PPP_g)^\real \cong M^g.$$

Applying the stationary phase approximation (see \cite[Formula
2.2]{AtiyahAsterisqueCircularSymm}) to the integral \eqref{path integral}, we
get
\begin{eqnarray}
\int_{(\PPP_g)^\real} \frac{e^{-t E(\phi)}}{\prod_j (tm_j -i\alpha_j)} =
\int_{M^g} \frac{1}{\prod_j (tm_j -i\alpha_j)}, \label{denominator}
\end{eqnarray}
where the energy of the constant paths is zero, the $m_j$ are
rotation numbers normal to $M^g$, and the $\alpha_j$ are the Chern
roots, so that the total Chern class of the normal bundle $N$ to
$M^g$ is given by
$$\prod_j (1 + \alpha_j).$$

\subsection{Normal bundle}

 For $f \in \PPP_g$, the tangent space $T_f$ at $f$ can be seen as the space of maps
$$\sigma : \real \to f^*TM$$
such that $\sigma(t) dg_{\sigma(t)} = \sigma (2\pi + t)$, so for
the constant map at $x \in M^g$, its tangent space is equal to the
space of maps
$$\sigma : \real \to T_xM$$
with $\sigma(t) dg_x = \sigma(2\pi + t)$.

We can split the vector space $T_xM$ into subspaces $N(\theta)$
that consist of 2-dimensional spaces on which $dg_x$ rotates every
vector by $\theta$ (see \cite{Lawson}):
$$T_xM \cong N(0) \oplus \bigoplus _\theta N(\theta).$$
It is clear that the number of $\theta$ is finite, that we could
choose them in the interval\footnote{For simplicity we will assume
that the eigenvalue $\pi$ is not included, in order to avoid
the use of Pontrjagin classes. The result still holds with $\pi$
as rotation number.} $0 < \theta < \pi$,
 and that $N(0) \cong T_xM^g$.

The constant functions
$$\{\sigma : \real \to T_xM^g \cong N(0) \mid \  \sigma \  \hbox{{\rm is constant}}\} \subset T_x \PP_g$$ give
the directions along $M^g$. We are interested in finding a
description of the normal directions of $M^g$ in $T_x\PPP_g$.

Let $2s(\theta) := {\rm dim}_\real N(\theta)$ and, for $l
= 1, \dots, s(\theta)$, let $N_l(\theta)$ be the 2-dimensional subspaces  fixed by
$dg_x$ through the rotation of $\theta$. Then any $\sigma \in T_x
\PPP_g$ can be seen as
$$\sigma  = \sum_{l,\theta} \sigma_l^\theta \ \ \ \ \ \ \
 {\rm with} \ \ \ \ \  \sigma_l^\theta : \real \to N_l(\theta).$$
Let $N_l(\theta)^\complex$ be the complexification $N_l(\theta) \otimes \complex$. Then
$$N_l(\theta)^\complex \cong L_l \oplus \overline{L_l},$$
where $L_l$ is a complex line bundle, the action of $dg_x$ on $L_l$
is by multiplication by $e^{i \theta}$, and $\overline{L_l}$ is the
conjugate bundle of $L_l$ (see \cite[p. 226]{Lawson}). The map
$$\sigma_l^\theta : \real \to N_l(\theta) \subset
N_l(\theta)^\complex$$ can be seen in $ L_l \oplus \overline{L_l}$
via a Fourier expansion as
\begin{eqnarray}
\sigma_l^\theta(t) = \sum_{k \in \integer} \left(
\begin{array}{c}
a_k\\
b_k
\end{array} \right)
\left(
\begin{array}{cc}
e^{itk}e^{it\frac{\theta}{2\pi}} & 0\\
0 & e^{itk}e^{-it\frac{\theta}{2\pi}}
\end{array} \right) \label{fourier}
\end{eqnarray}
with $a_k \in L_l$, $b_k \in \overline{L_l}$, $a_k =
\overline{a_{-k}}$ and $b_k = \overline{b_{-k}}$ (the last two
equations hold because $\sum_k a_k e^{itk}$ and $\sum_k b_k
e^{itk}$ are real for all $t$; in particular $a_0, b_0 \in
\real$).

Then the tangent bundle to $T_x \PPP_g$ can be decomposed as an
infinite direct sum
$$T_xM \oplus (T_xM^\complex)_1 \oplus (T_xM^\complex)_2 \oplus \cdots$$
with $$(T_xM^\complex)_n \cong (N(0)^\complex)_n \oplus
\bigoplus_\theta (N(\theta)^\complex)_n$$ where the circle acts in
each $(N(\theta)^\complex )_n$ by rotation number $n$. The
coefficients $(a_k,b_k)$ of the Fourier expansion of
\eqref{fourier} take values in $(N(\theta)^\complex)_k$ for $k>0$,
$(a_0,b_0) \in N(\theta)$, and  $(a_k,b_k) =
(\overline{a_{-k}},\overline{b_{-k}})$ for $k<0$.

As $T_x M \cong N(0) \oplus \bigoplus_\theta N(\theta)$ and $N(0)
\cong T_xM^g$ represent the directions along $M^g$, the normal
bundle to $M^g$ in $\PPP_g$ can be represented as
$$\left\{ (N(0)^\complex)_1 \oplus (N(0)^\complex)_2 \oplus \cdots \right\} \oplus
\bigoplus_\theta \left\{  N(\theta) \oplus (N(\theta)^\complex)_1
\oplus (N(\theta)^\complex)_2 \oplus \cdots \right\}.$$
Let the
Chern class of $N(\theta)$ be
$$\prod_{k=1}^{s(\theta)} (1 + y_k^\theta),$$
so its $g$-Chern character is
$$ch_g(N(\theta)) = \sum_{k=1}^{s(\theta)} ch(N_k(\theta))\chi(g) = \sum_{k=1}^{s(\theta)} e^{y_k^\theta + i \theta};$$
 then the $g$-Chern class of the complexification of $N(\theta)$ is
$$\prod_{k=1}^{s(\theta)} (1+ y_k^\theta +i\theta)(1-y_k^\theta - i \theta).$$

If we let $x_k$ denote the Chern classes of $M^g$, then the
denominator in \eqref{denominator} with $t=1$ becomes
\begin{eqnarray*}
\prod_{j=1}^{s(0)} \prod_{p=1}^\infty \left( p^2 + x_j^2 \right) \prod_\theta \left\{ \prod_{k=1}^{s(\theta)} (y_k^\theta + i \theta)
\prod_{p=1}^\infty \left(p^2 + (y_k^\theta + i \theta) \right) \right\},
\end{eqnarray*}
which is formally
\begin{eqnarray*}
\prod_{j=1}^{s(0)} \left( \prod_{p=1}^\infty p^2 \right) \frac{\sinh(\pi x_j)}{\pi x_j}
 \prod_\theta \left\{ \prod_{k=1}^{s(\theta)}
\left( \prod_{p=1}^\infty p^2 \right)
(y_k^\theta + i \theta) \frac{\sinh(\pi(y_k^\theta + i \theta))}{\pi(y_k^\theta + i \theta)} \right\}.
\end{eqnarray*}
Replacing the infinite product of the $p^2$ by its renormalized factor $2\pi$, we get
\begin{eqnarray*}
\prod_j \frac{2\sinh(\pi x_j)}{x_j} \prod_\theta \left\{ \prod_k 2\sinh(\pi (y_k^\theta + i \theta)) \right\},
\end{eqnarray*}
which is the same as
\begin{eqnarray}
\prod_j \frac{\sinh(x_j / 2)}{x_j /2} \prod_\theta \left\{ \prod_k \frac{\sinh((y_k^\theta + i \theta)/2)}{1/2} \right\}
\label{productofsinh}
\end{eqnarray}
provided we interpret $\prod_{p=1}^\infty t$ as $t^{\zeta(0)}$
where $\zeta(s)$ is the Riemann zeta function. As $\zeta(0) =
-\frac{1}{2}$, in each component we get a factor of $t$ which
cancels with the factor $t^{-1}$ that arises from replacing $x_j$
by $x_j/t$ and $y_k^\theta + i \theta$ by $(y_k^\theta + i
\theta)/t$. Our use of the stationary phase approximation is
independent of $t$, and setting $t=2\pi$ we get formula
\eqref{productofsinh}.

In the notation of \cite[p. 267]{Lawson} formula \eqref{productofsinh} is equivalent to
\begin{eqnarray*}
\left( \hat{A} (M^g) \prod_\theta \hat{A}(N(\theta)) \right)^{-1},
\end{eqnarray*}
which after replacing it in the denominator of \eqref{denominator}
and integrating over $M^g$
 matches the formula for $Spin(M,g)$ \cite[Th. 14.11]{Lawson}:
\begin{eqnarray*}
Spin(M,g) = (-1)^{\tau_g} \hat{A}(M^g) \left\{\prod_\theta \hat{A}(N(\theta))  \right\} [M^g].
\end{eqnarray*}

We conclude that after applying the stationary phase
approximation to \eqref{path integral}, we obtain the $g$-index of
the Dirac operator.

\begin{proposition}
The path integral
\begin{eqnarray*}
\int_{\PPP_g} e^{-t E(\phi)} \{{\rm Tr} \ S^+(T_\phi) -{\rm Tr} \ S^-(T_\phi)\} d\phi = Spin(M,g)
\end{eqnarray*}
equals ${\rm ind}_g(D^+)$, the g-index of the Dirac operator over $M$.
\end{proposition}

\subsection{The $G$-index and Kawasaki's formula}
The $G$-index of the Dirac operator is an element of $R(G)$, the
representation ring of $G$. Using localization, its dimension is
equal to
\begin{eqnarray*}
{\rm ind}_G (D^+)=\frac{1}{|G|} \sum_{g \in G} {\rm ind}_g (D^+) =\frac{1}{|G|} \sum_{g \in G} Spin(M,g).
 \end{eqnarray*}
But instead of summing over all the elements $g$ in $G$, we could
sum over the conjugacy classes of $G$. It is clear that $Spin(M,g)
= Spin(M,h^{-1}gh)$.  The size of the conjugacy class $(g)$ of $g$
is $\frac{|G|}{|C(g)|}$ where $C(g)$ is the centralizer of $G$, that is,
the set of elements which commute with $g$ (equivalently,
the fixed point set
of the action of $G$ in $g$ via conjugation). Thus, we obtain

\begin{eqnarray*}
{\rm ind}_G(D^+) = \sum_{(g)} \frac{1}{|C(g)|} Spin(M,g).
\end{eqnarray*}

We would like to derive a formula that depends on the twisted
sectors (inertia groupoid) of the orbifold $\Xx = [M/G]$, and this
clearly matches our previous description. In
\cite{LupercioUribeLoopGroupoid} it was argued that the fixed
point set of the action of $\real$ in the loop groupoid $\Loop
\Xx$ was precisely $I( \Xx)$ the inertia groupoid of $\Xx$; then,
applying stationary phase approximation
 to
\begin{eqnarray*}
\int_{\Loop \Xx} e^{-t E(\phi)} \{{\rm Tr} \ S^+(T_\phi) -{\rm Tr} \ S^-(T_\phi)\} \DD\phi,
\end{eqnarray*}
which can be rewritten as
\begin{eqnarray*}
\sum_{(g)} \frac{1}{|C(g)|} \int_{\PPP_g} e^{-t E(\phi)} \{{\rm Tr} \ S^+(T_\phi) -{\rm Tr} \ S^-(T_\phi)\} \DD\phi,
\end{eqnarray*}
we get the $G$-index of the Dirac operator,
\begin{eqnarray*}
{\rm ind}_G(D^+)=
\sum_{(g)} \frac{(-1)^{\tau_g}}{|C(g)|} \int_{M^g}  \hat{A} (M^g) \prod_\theta \hat{A}(N(\theta)_g).
\end{eqnarray*}

%
Which can be shown to coincide with the formula given by Kawasaki \cite[p.
139]{Kawasaki} for the index theorem for $V$-manifolds. Thus,
 the localization principle applies in this case.

\section{The Elliptic Genus}

We move on now to localizing functional integrals in the double
loopspace $\LL^2 M = \LL \LL M = \Map(T,\Map(T,M))=\Map(T^2,M),$
where $T=S^1$ and $T^2$ is the 2-torus. By performing the
corresponding functional integral over $\LL^2 M$, we should obtain
the index of the dirac operator over $\LL M$ considered by Witten
in \cite{WittenElliptic} and known as the elliptic genus
\cite{SegalAsterisqueElliptic}. This has been verified by Ando and
Morava \cite{AndoMorava}. We want to perform the calculation in
the orbifold case (cf. \cite{AndoFrench}).

Let the groupoid $\Xx$ be $[M/G]$, and let the torus $\Tt$ be
represented by the groupoid $[\real^2/\integer \oplus \integer]$.
The {\it double loop groupoid} $\Loop^2 \Xx$ is the category with
smooth functors $\Tt \to \Xx$ as objects and
natural transformations between functors as morphisms.

A morphism in $\Loop^2\Xx$ can be seen as

$$ \begin{array}{ccc}
\real^2 \times (\integer \oplus \integer)  & \longrightarrow & M \times G\\
\downdownarrows & & \downdownarrows \\
\real^2& \longrightarrow & M,
\end{array}
$$
that is, as a map $F \colon \real^2 \to M$ together with a
homomorphism $H \colon \integer \oplus \integer \to G$ such that $F$ is
equivariant with respect to $H$. This is equivalent to choosing a
pair of commuting elements $g , h \in G$ such that $F(1,0) =
F(0,0)g$, $F(0,1)= F(0,0)h$ and in general $F(n,m) = F(0,0)
g^nh^m$.

\begin{center}
\includegraphics[width= 3.5 in, height= 1.4 in]{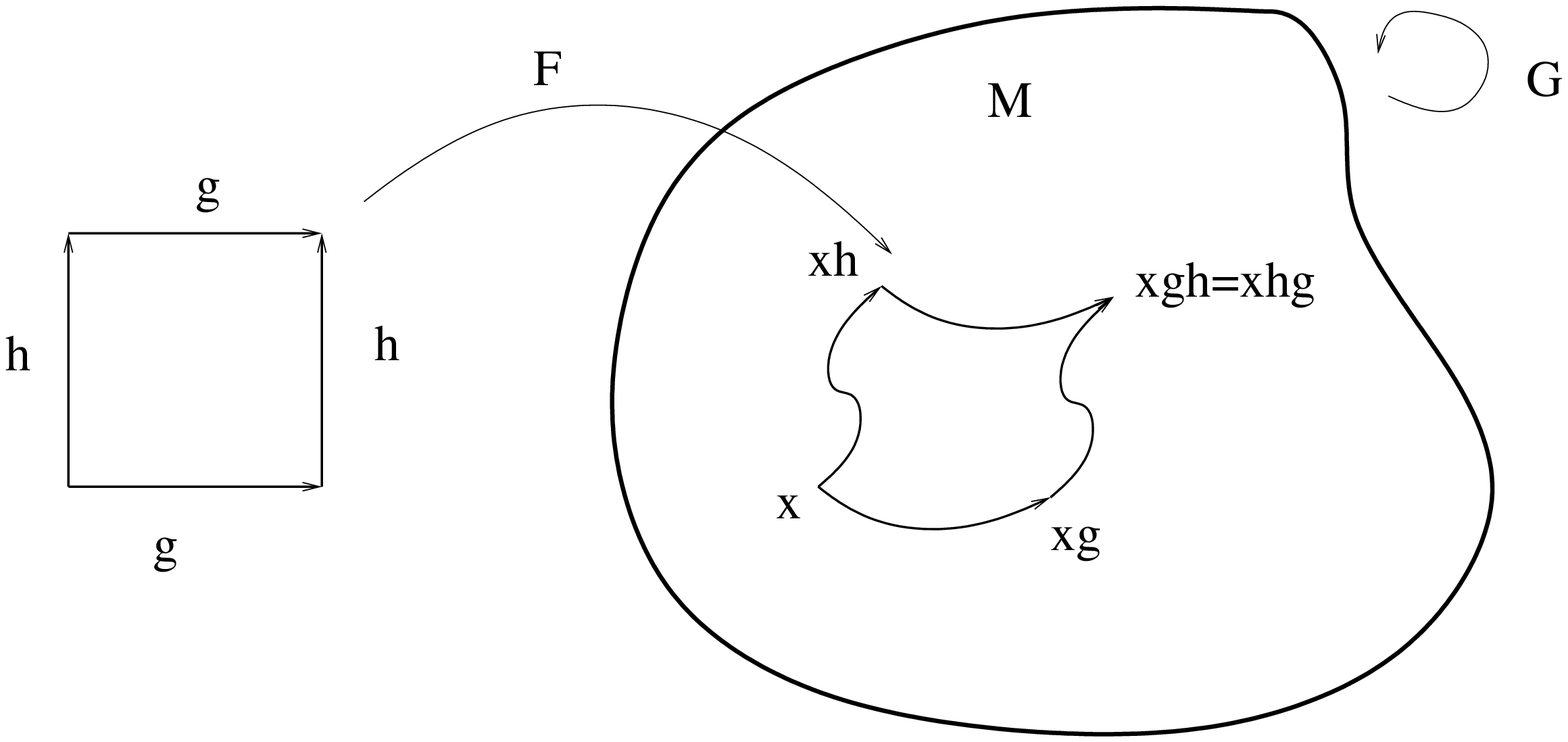}
\end{center}

The group $\real^2$ acts naturally by translations on the double
loop groupoid. This action factors through $\real^2/\{
|G|\integer \oplus |G|\integer \}$ because every orbifold loop can
be closed in $M/G$.

The fixed points under the action of $\real^2$ are the
constant double loops; they are uniquely determined by a choice
of a point in $M$ and two commuting elements in $G$.

The groupoid of {\it{ghost double loops}} is the groupoid whose
objects is the set of functors
$$Funct( [* / \integer \oplus \integer] , [M/G])$$
and whose morphisms are natural transformations (i.e., it is a
groupoid $[(Funct( [* / \integer \oplus \integer] , [M/G])) /G ]$ with  $G$ acting by conjugation on the
functors).

Here we will apply the stationary phase approximation  formula to
the double loop groupoid, which we have shown above to be endowed
with an action of the torus.

We will use an alternative description of the double loop groupoid.
Its elements will be smooth maps
$$\phi \colon [0,1]^2 \to M$$
together with commuting elements $g,h \in G$ such that $\phi(1,0)
= \phi(0,0)g$, $\phi(0,1)= \phi(0,0)h$. Call this set
$\LL^2_{\langle g,h \rangle} M$ and take
$$\LL^2 M := \bigsqcup_{\{(g,h) \in G^2 |gh =hg \}} \LL^2_{\langle g,h \rangle} M.$$
 The natural action of conjugation by elements in $G$ gives us the description:
$\Loop ^2 \Xx \cong [(\LL^2 M)/G]$.

We consider the functional of double loops

$$\HH(\phi) := \int_{[0,1]^2} (|| \frac{d\phi}{ds} ||^2 + ||\frac{d\phi}{dt} ||^2) ds\, dt;$$
we will apply stationary phase approximation \`{a} la
Witten-Atiyah to the Feynman integral
$$\int_{\Loop^2 \Xx} e^{-i \HH(\phi) } \DD \phi.$$
We need to find the equivariant normal bundle on $\Loop^2 \Xx$ to the
fixed points of the action of $\real^2$, namely the ghost double loops.

For commuting $g,h \in G$, take the part of the groupoid of
ghost double loops
 parameterized by $M^{\langle g,h \rangle}$, the fixed point
set of the group generated by $g$ and $h$. Call $\iota :
M^{\langle g, h \rangle} \hookrightarrow M$ the  inclusion, and
 suppose the the orbifold $\Xx$ is a complex
orbifold (the pullback bundle $\iota^* TM$ can be locally
simultaneously diagonalized with respect to the actions of $g$ and
$h$). Then one can write the total Chern class of $\iota^* TM$ as $
\prod_j (1 + x_j)$ such that the line bundle $x_j$ comes provided
with the action of the group $\langle g,h \rangle $ parameterized
by the irreducible representation $\lambda_j$.

We are using the following fact about equivariant complex $K$-theory.
If a group $\Gamma$
acts trivially on a space $Y$, then
$$K^*_\Gamma (Y) \cong K^*(Y) \otimes R(\Gamma),$$
that is,
the equivariant $K$-theory of $Y$ is isomorphic to the ordinary
$K$-theory of $Y$
tensored with the representation ring of $\Gamma$.
Then the equivariant Chern character associated to the $\langle
g,h\rangle$ equivariant line bundle $x_j$ is $ch_{\langle
g,h\rangle}(x_j) = e^{x_j} \otimes \chi_{\lambda_j}$, where
$e^{x_j}$ is the Chern character of the line bundle and
$\chi_{\lambda_j}$ is the character of the $\lambda_i$
representation.
As we have simultaneously diagonalized the actions of $g$ and $h$,
the character of an irreducible representation is determined
by a root of unity associated to each $g$ and $h$. So let
$\sigma_j \colon \langle g, h \rangle \to [0,1)$ be such that
$\chi_{\lambda_j}(g) = e^{2 \pi i \sigma_j(g)}$; then one can
consider $(1+ x_j +  2 \pi i \sigma_j)$ as the Chern class of the
equivariant bundle $x_j$.

The equivariant Euler class of the normal bundle of the embedding of ghost double loops
$$M^{\langle g,h \rangle} \to \LL^2_{\langle g,h \rangle} M$$
is then
$$\left\{\prod_{\{ j | \sigma_j(g)=\sigma_j(h)=0\}}\frac{1}{x_j} \right\}
\left \{ \prod_j \prod _{(k,l) \in \integer^2}  \left( x_j +
l\hat{p}  + k \hat{q} +  \sigma_j(g)\hat{p} + \sigma_j(h)\hat{q}
\right) \right\},$$
where $\hat{p}$ and $\hat{q}$ are formal
variables that keep track of the fractional periods of each of the
circles of the torus.

Applying the fixed point formula $(3.2.1)$ of Ando-Morava
\cite{AndoMorava}, one obtains
$$p^{\LL^2_{\langle g,h \rangle} M}(1) = p^{M^{\langle g,h \rangle}} \left\{\prod_{\{ j | \sigma_j(g)=\sigma_j(h)=0\}} x_j \right\}
\left \{ \prod_j \prod _{(k,l) \in \integer^2}  \frac{1}{ x_j + l\hat{p}  + k \hat{q} +\sigma_j(g)\hat{p} + \sigma_j(h)\hat{q} }
\right\}.
$$
Rearranging the expression in the second parenthesis by factoring
$k\hat{q}$ and keeping the $l$ fixed,  the second
parenthesis becomes:
$$ \prod_{l \in \integer} \left(\prod_{k >0} \frac{1}{k^2 \hat{q}^2} \right) \left( (x_j + l\hat{p} +  \sigma_j(g)\hat{p} +  \sigma_j(h)\hat{q})
\prod_{k>0} \left(1 - \frac{(x_j + l\hat{p}
+  \sigma_j(g)\hat{p} + \sigma_j(h)\hat{q})^2}{k^2 \hat{q}^2} \right) \right)^{-1}.$$

Renormalization
(see \cite{AtiyahAsterisqueCircularSymm, AndoMorava}) gives
$$\prod_{k >0} \frac{1}{k^2 \hat{q}^2} = \frac{\hat{q}}{2 \pi},$$
$$\prod_{k>0} \left(1 - \frac{(x_j + l\hat{p}
+ \sigma_j(g)\hat{p} + \sigma_j(h)\hat{q})^2}{k^2 \hat{q}^2}
\right)^{-1} =\frac{\hat{q}}{2 \pi} \frac{
\frac{\pi}{\hat{q}}}{\sin\left( \frac{\pi}{\hat{q}} (x_j +
l\hat{p} +  \sigma_j(g)\hat{p} + \sigma_j(h) \hat{q})\right)}.$$
Replacing the variable $\hat{q}$ by its holonomy $2 \pi i$, our
push forward $p^{\LL^2_{\langle g,h \rangle} M}(1)$ becomes
$$p^{M^{\langle g,h \rangle}} \left\{\prod_{\{ j | \sigma_j(g)=\sigma_j(h)=0\}} x_j \right\} \left\{
\prod_j \prod_{l \in \integer} \frac{\frac{1}{2}}{\sinh \frac{1}{2}(x_j + l\hat{p} + \sigma_j(g)\hat{p} + 2 \pi i \sigma_j(h)) } \right\};$$
pairing the hyperbolic sines of $l$ and $-l$ one gets that
$$2 \sinh \frac{(x_j + l\hat{p} +\sigma_j(g)\hat{p}+2 \pi i \sigma_j(h))}{2} 2 \sinh \frac{(x_j - l\hat{p} +\sigma_j(g)\hat{p}
+2 \pi i \sigma_j(h))}{2} = \ \ \ \ \ \ \ \ \ \ \ \ \ \ \ \ \ \ \ \ \ \ \ $$
$$\ \ \ \ \ \ \ \ \ \ \ \ \ \ \ \
 \frac{1-e^{-x_j  -\sigma_j(g)\hat{p} -2 \pi i \sigma_j(h) - l\hat{p}} }{e^{-\frac{1}{2}(x_j +\sigma_j(g)\hat{p}+ 2\pi i \sigma_j(h))} e^{-\frac{l}{2} \hat{p}}}
\frac{e^{x_j  +\sigma_j(g)\hat{p}+2 \pi i \sigma_j(h) - l\hat{p} }-1 }{e^{\frac{1}{2}(x_j +\sigma_j(g)\hat{p}+ 2\pi i \sigma_j(h))} e^{-\frac{l}{2} \hat{p}}}.$$
As a result,
$$p^{\LL^2_{\langle g,h \rangle} M}(1) =
p^{M^{\langle g,h \rangle}} \left\{\prod_{\{ j | \sigma_j(g)=\sigma_j(h)=0\}} x_j \right\} \times \ \ \ \ \ \ \ \ \ \ \  \ \ \ \ \ \ \ \ \ \ \ \ \ \ \ $$
$$ \left\{
\prod_j \frac{\frac{1}{2}}{\sinh\frac{1}{2}(x_j + \sigma_j(g)\hat{p} +2\pi i \sigma_j(h))}
 \prod_{l >0} \frac{ -e^{-\hat{p}l} }{(1- e^{-x_j  -\sigma_j(g)\hat{p}-2 \pi i \sigma_j(h) - l\hat{p} })(1-e^{x_j +\sigma_j(g)\hat{p} +2 \pi i \sigma_j(h) - l\hat{p}} )} \right \}
$$
$$=
p^{M^{\langle g,h \rangle}} \left\{\prod_{\{ j | \sigma_j(g)=\sigma_j(h)=0\}} x_j \right\} (-e^{\hat{p}})^{\frac{1}{12}} \times\ \ \ \ \ \ \ \ \ \ \  \ \ \ \ \ \ \ \ \ \ \ \ \ \ \ $$
$$\left\{
\prod_{l >0,j} \frac{ e^{\frac{1}{2}(-x_j  -\sigma_j(g)\hat{p}-2 \pi i \sigma_j(h))} }{(1- e^{-x_j  -\sigma_j(g)\hat{p}-2 \pi i \sigma_j(h) -(l-1)\hat{p} })(1-e^{x_j +\sigma_j(g)\hat{p} +2 \pi i \sigma_j(h) - l\hat{p}} )} \right \}.
$$
Making the change of variables $p=e^{-\hat{p}}$, assuming that the first Chern
class of $M$ satisfies $c_1(M)=0$, i.e. $\prod_j e^{x_j} =1$, and
integrating over $M^{\langle g,h \rangle}$, we have that
$$p^{\LL^2_{\langle g,h \rangle} M}(1) = \frac{p^{\left(-\frac{dim(M)}{12} +i \pi+ \frac{\age{g}}{2}\right) } e^{- \pi i \age{h}} \left\{ \prod_{\{ j | \sigma_j(g)=\sigma_j(h)=0\}} x_j \right\} }
{
\prod_{l >0,j} {(1- p^{l-1 +\sigma_j(g)}e^{-x_j  -2 \pi i \sigma_j(h) })(1-p^{l - \sigma_j(g)}e^{x_j +2 \pi i \sigma_j(h)  }})} [M^{\langle g,h \rangle}].
$$

\vspace{1cm}
 Adding all the fixed point data and averaging,
one gets the orbifold elliptic genus:
$$Ell_{orb}([M/G]) = \ \ \ \ \ \ \ \ \ \ \ \ \ \ \ \ \ \ \ \ \ \ \ \ \ \ \ \ \ \ \ \ \ \ \ \ \ \ \ \ \ \ \ \ \ \ \ \ \ \ \ \ \ $$
$$ \frac{1}{|G|} \sum_{gh=hg}
 \frac{p^{\left(-\frac{dim(M)}{12} +i \pi+ \frac{\age{g}}{2}\right) } e^{- \pi i \age{h}} \left\{ \prod_{\{ j | \sigma_j(g)=\sigma_j(h)=0\}} x_j \right\} }
{
\prod_{l >0,j} {(1- p^{l-1 +\sigma_j(g)}e^{-x_j  -2 \pi i \sigma_j(h) })(1-p^{l - \sigma_j(g)}e^{x_j +2 \pi i \sigma_j(h)  }})} [M^{\langle g,h \rangle}].
$$

This coincides with the constant term in the $y$-expansion of the
formula of Borisov-Libgober \cite{BorisovLibgober,
DijkgraafMooreVerlindeVerlinde1997, DongLiuMa2002} except for a
renormalization factor. One could use a device like that of
Hirzebruch \cite{Hirzebruch} to recover the full formula. In any
case the localization principle holds in this case.

\bibliographystyle{amsplain}
\bibliography{Localizaton_Orbifold}

\end{document}